# Impact of mobility models on clustering based routing protocols in mobile WSNs

Atta ur Rehman Khan[1], Shahzad Ali[2, 3], Saad Mustafa[3], Mazliza Othman[1]

[1] Faculty of Science and Information Technology, University of Malaya,
Kuala Lumpur, Malaysia
[2] IMDEA Networks,
Avenida del Mar Mediterraneo, Leganes, Madrid
[3] Department of Computer Science, COMSATS Institute of Information Technology,
Abbottabad, Pakistan

*attaurrehman@siswa.um.edu.my, shahzadali@ciit.net.pk, saadmustafa@ciit.net.pk, mazliza@um.edu.my*

*Abstract—* *This paper presents comparison of different hierarchical (position and non-position based) protocols with respect to different mobility models. Previous work mainly focuses on static networks or at most a single mobility model. Using only one mobility model may not predict the behavior of routing protocol accurately. Simulation results show that mobility has large impact on the behavior of WSN routing protocols. Also, position based routing protocols performs better in terms of packet delivery compared to non position based routing protocols.*

## I. INTRODUCTION

Wireless sensor network (WSN) is an emerging class of ad hoc networks. WSNs comprise sensor nodes distributed over geographic area to monitor certain phenomenon. The sink node (base station) acts as gateway and is comparatively resourceful [1], whereas, sensor nodes have constrained energy, processing capacity, and memory [2].

Previous studies mostly consider evaluation based on static networks [1-3]. There are various applications where nodes are mobile and needs due consideration [1]. Mobile sensor network applications include battlefield surveillance, habitat monitoring, and search and rescue operations. Routing in mobile WSNs becomes more difficult because of the frequent path failures and unpredictable topology changes, which may increase packet loss and packet delay. Different routing protocols exist that can be broadly classified into hierarchical and flat routing. Some of the flat routing protocols are not scalable due to the assumption that sensor nodes can directly send data to the sink node [4-6]. Therefore, hierarchical routing protocols [7, 8-9] are preferred if scalability is the deciding factor or the number of nodes is very high.

When nodes are mobile, the performance of hierarchical (clustering based) protocols suffers due to two main reasons i.e., path breakage and consequently packet loss during intra cluster and inter cluster communication. Protocols having no backup strategy to deal with such situation yield low packet delivery ratio. Mostly the existing comparative studies consider only one mobility model for evaluating routing protocols. Only one mobility model does not reflect the true behavior of a protocol [10], therefore we have tested the selected protocols with three different mobility models.

In our work, we have selected a few position and non-position based hierarchical clustering protocols. All the selected protocols are studied thoroughly for several speeds, ranging from low to high speed movement with respect to different mobility models. By extensive simulations, it is concluded that position-based routing protocols performs better than non position based routing protocols in terms of packet delivery ratio.

The paper is organized in five sections. Section 2 presents the related work. Section 3 explains the comparison strategy. Section 4 presents the simulation results, and Section 5 concludes the paper.

## II. RELATED WORK

One of the most important design goals of WSNs is to minimizing energy consumption of the network. Moreover, if sensor nodes are mobile, it further complicates the design of the network. To comprehensively simulate a newly proposed protocol for mobile sensor networks, it is recommended to check the performance of the protocol with multiple mobility models [11]. This is because performance of every protocol is dependent on the application scenario.

A given protocol can perform well in one environment and can fail when environment is changed [12]. Same is the case with mobility models. A given protocol can perform well for one mobility model but can exhibit deteriorated performance under some other mobility model. This is because the performance of mobile WSN routing protocols is highly affected by the mobility models. In [11], authors have shown that mobility models can affect packet delay, packet delivery ratio, and control overhead of a routing protocol.

In [13], authors highlighted the importance of underlying mobility models and simulated the results for different mobility models. They also discussed seven synthetic entity mobility models and five group mobility models. In this, authors concluded from simulation results that Random Way Point Mobility Model has highest packet delivery ratio and lowest end to end delay compared to other selected mobility models.

In [14] authors investigated impact of mobility models on performance of specific network protocol or application and different routing protocols were evaluated under different mobility patterns. Simulation results show that different mobility patterns affect routing protocols in different ways. They also concluded that selection of mobility model alters physical link dynamics and cluster stability. So ranking of routing protocols is dependent on the selection of the mobility pattern.

In [15], the authors simulated same protocol for different mobility models and concluded that performance of protocol is not only affected by different mobility models but also by different parameters of same mobility model. Moreover, a routing protocol should be simulated for mobility model that closely resemblances its real world application.

Hierarchical routing protocols are extensively tested for ad hoc networks [6, 14, 16-25]. In hierarchical routing protocols, the main process is cluster formation. For cluster formation, few nodes are selected as cluster heads and remaining nodes associate themselves with the cluster heads. Nodes along with associated cluster head is known is cluster. Nodes are responsible to sense information from surrounding and send that information to the associated cluster head. This process is called intra cluster communication. When information is received by cluster head it performs different operations and finally data is routed to other cluster heads, the phenomenon known as inter cluster communication.

Hierarchical based clustering protocols can be further divided into position based and non-position based protocols routing protocols. In our work we have selected few position based and non-position based routing protocols. Each of these protocols with respect to their categories is explained below.

*A. Position Based Protocols*

Position based routing protocols are dependent on the location information. Location of sensor nodes can be identified with the help of low power GPS module embedded in sensor nodes or some distributed localization technique [16, 19-20, 21-22]. By using location information, many tasks can be done efficiently [23]. For example, based on the distance between two nodes, energy consumption can be estimated for all routing paths between the two nodes and then select more energy efficient path [24]. Position based routing protocols, that are considered during our work are Mobility Aware Routing Protocol (MAR) [25] and Distributed Geographic Clustering Protocol (GRC) [26].

*1) Mobility Aware Routing Protocol (MAR)*

In MAR, cluster heads are selected on the basis of mobility. Nodes which are less mobile are selected as cluster heads and this mechanism of cluster head selection leads to more stable clusters. MAR does not consider residual energy of nodes during selection of cluster heads hence are energy unaware. Moreover, due to mobility of sensor nodes, cluster heads may move out of transmission range of each other. Also MAR does not have any packet recovery mechanism for inter-cluster communication and is location unaware, so packet loss occurs.

*2) Distributed Geographic Robust Clustering Protocol (GRC)*

GRC is energy aware routing protocol and uses location information for selection of cluster head. Those nodes which are more close to center of specified zones and have higher residual energy are selected as cluster heads. The purpose of introducing "center-ness" factor in selection of cluster heads was to make sure that even if there is mobility, the head will take some time to have substantial movement and get out of the range of the cluster nodes.

To minimize packet loss during inter-cluster communication a recovery mechanism was introduced in GRC. Two versions of GRC are used during simulations which are GRC without recovery strategy and GRC with recovery strategy (GRC-R).

*B. Non Position Based Protocols*

Non position based routing protocols do not need any position information to make their routing decisions. Non-position based routing protocols, that we have selected includes Distributed Efficient Clustering Approach (DECA) [17] and Distributed Efficient Multi hop Clustering protocol (DEMC) [27].

*1) Distributed Efficient Clustering Approach (DECA)*

DECA is a non position-based protocol that considers node mobility, node residual energy, identifier, and its connectivity with other nodes; all these parameters are used to calculate weight for each node. Only one message is transmitted during clustering, which saves more energy as compared to low energy adaptive cluster hierarchy (LEACH) [5] and Hybrid Energy Efficient Distributed Clustering (HEED) [7] which sends multiple clustering messages during clustering phase As transmission and reception are main sources of energy consumption in sensor networks [28], so by reducing number of messages; DECA becomes more energy efficient. But the major problem with DECA is that it does not use any recovery mechanism for inter-cluster communication resulting in packet loss.

### 2) Distributed Efficient Multi hop Clustering protocol (DEMC)

DEMC is a distributed clustering based routing protocol specially designed for mobile sensor networks and is more energy efficient compared to DECA. This is because DEMC does not send periodic hello messages, does not keep neighbors list and requires only one message per cluster for selection of cluster head. By removing extra overhead and minimizing control messages, DEMC is more energy efficient in comparison to DECA. To minimize inter-cluster communication packet loss, a recovery mechanism was also introduced in DEMC. Therefore, DEMC also incurs less packet loss compared to DECA. Two versions of DEMC were used during simulation. One is simple DEMC without recovery strategy and the other is DEMC with recovery strategy.

## III. COMPARISON STRATEGY

Our focus is on comparison of position and non position based clustering protocols with respect to different mobility models. In this section, we discuss the performance metrics that are used for comparative study as well as the mobility models used.

### A. Performance Metrics

For evaluating performance of selected protocols, following metrics were used.

#### 1) Percentage of Packet loss

With this performance metric we can check reliability of a protocol, the lower the percentage of packet loss the more reliable it is. Let 'n' be the total packets sent and 'm' be the received packets then percentage packet loss is calculated as

$$((n-m)/n) \times 100. \qquad (1)$$

We can calculate percentage of packet loss by dividing total number of lost packets by total number of sent packets. When comparing multiple protocols, the one which has lower percentage for packet loss is considered better compared to those which have high percentage of packet loss.

#### 2) Packet delivery ratio

This is also one of the good performance metrics and is calculated by dividing total number of delivered packets by total number of sent packets. For any good protocol the ideal packet deliver ratio must be one. For some protocols packet delivery ratio can be higher than one which is also not a good sign because in that case packet duplication is occurring. So, one must try to achieve packet delivery ratio closer to one but not greater than one. The closer the packet delivery ratio to one, the better the protocol performance is.

### B. Mobility Models

The following three mobility models were used during simulations.

#### 1) Linear Mobility Model

In linear mobility model [29] nodes move in a straight line with a certain angle and this angle changes only when the mobile node hits a wall: then it reflects off the wall at the same angle.

#### 2) Mass Mobility Model

Mass mobility model [29] is a variant of random waypoint mobility model. In this mobility model nodes are considered to be having some mass and then apply momentum accordingly. Due to this factor, nodes do not turn, starts, or stops instantaneously.

#### 3) Random Way Point Mobility Model

In random way point mobility model, nodes move with random speed towards randomly selected destination. As random waypoint mobility model does not consider mass of node so in this mobility model nodes can turn, start, and stop instantaneously.

## IV. SIMULATION AND RESULTS

All the simulations are done in OMNET++ based simulation framework called INET [29]. The reason for using this framework is that it is suitable for simulations of sensor networks and moreover it supports various mobility models as well [30].

For simulations, using uniform distribution, 100 nodes were distributed randomly in the network field with 1000m × 1000m dimensions. Then selected protocols are tested with different mobility models using different parameters for mobility and varying number of nodes.

Figures 1-6 shows packet loss percentage and packet delivery ratio for DECA, DEMC, DEMC-R, MAR, GRC and GRC-R. All these protocols were simulated with respect to different speeds using random waypoint mobility model, mass mobility model, and linear mobility model to investigate performance of these protocols in terms of packet delivery ratio and packet loss.

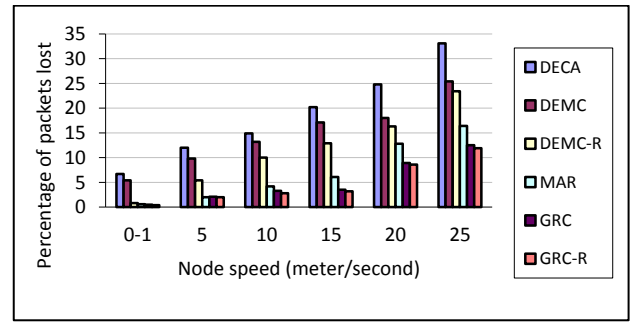

Fig.1 Percentage packet loss with random waypoint mobility model

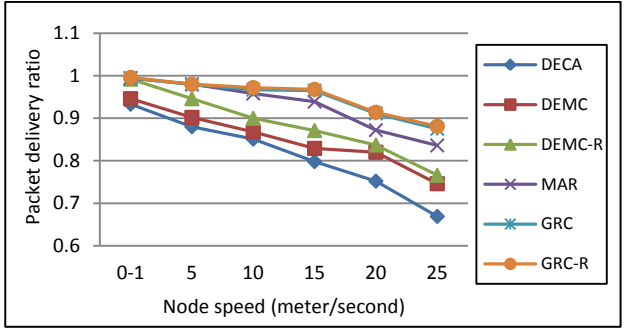

Fig.2 Packet delivery ratio with random waypoint mobility model

A prominent thing that is evident from the simulation results is that the protocols having recovery mechanism perform much better as compared to protocols having no such mechanism. The reason behind this is that, recovery mechanism in inter-cluster communication minimizes the packet loss and therefore increase packet delivery ratio.

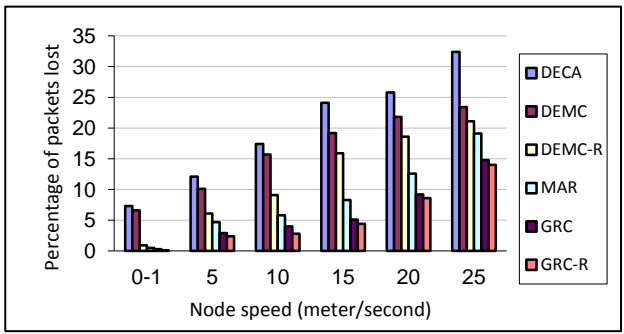

Fig.3 Percentage packet loss with respect to mass mobility model

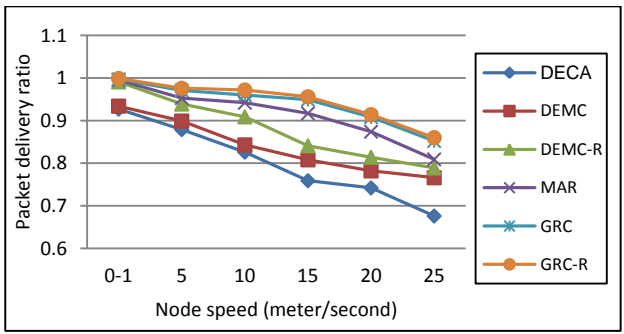

Fig.4 Packet delivery ratio with respect to mass mobility model

For the specified protocols mass mobility model incurs highest packet loss compared to random waypoint mobility model and linear mobility model. This is because in mass mobility model is a more realistic mobility model as compared to other two mobility models; and in mass mobility nodes take smooth turns and goes out of cluster-head range.

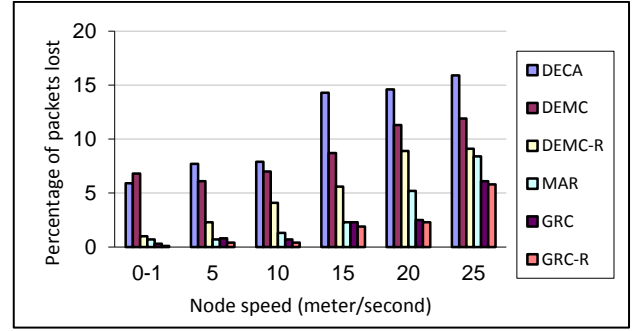

Fig.5 Percentage packet loss with respect to linear mobility model

In random waypoint mobility model, nodes stops at random locations and perform abrupt turns, during these turns, mostly the direction is toward cluster-head. For this reason packet loss of random waypoint mobility model is less compared to mass mobility model. With linear mobility model all protocols have shown minimal packet loss. This is because mobility of all nodes is not only associated with each other but also they move in a straight line until they hit some fixed object. After hitting any fixed point all nodes turn with specific angle and goes in some other direction. In this way nodes mostly move together, not away from each which leads to less packet loss and high packet delivery ratio.

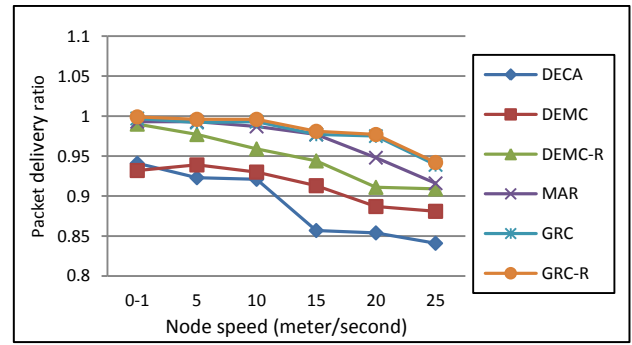

Fig.6 Packet delivery ratio with respect to linear mobility model

Simulation results have shown that non position based protocols (DECA, DEMC and DEMC-R) incur high packet loss and low packet delivery ratio compared to position based protocols (MAR, GRC and GRC-R). So no matter what mobility model is used, position based routing protocols always perform better than non position based routing protocols in terms of packet loss and packet delivery ratio.

V. CONCLUSION

In this paper we selected clustering based routing protocols from position based and non position based categories and compared their performance with respect to different mobility models. From simulation results, it is evident that the network performance is enhanced in the presence of a recovery mechanism. All the protocols were tested under different speeds using three different mobility models. Overall position-based routing protocols provide high packet delivery ratio due to less packet loss. On the other hand non position based routing protocols provide low packet delivery ratio and high packet loss compared to position based routing protocols. It would be interesting to see the overhead incurred by the inclusion of recovery mechanism in the protocols. Also, in future we would include more routing protocols and more mobility models and present results.

REFERENCES

[1] F. Tashtarian, A.T. Haghighat, M.T. Honary, H. Shokrzadeh, "A New Energy-Efficient Clustering Algorithm for Wireless Sensor Networks," 15th International Conference SoftCOM on Software, Telecommunications and Computer Networks, September 2007, pp.1 - 6.
[2] Q. Wang, W. Yang, "Energy Consumption Model for Power Management in Wireless Sensor Networks," 4th Annual IEEE Communications Society Conference on Sensor, Mesh and Ad Hoc Communications and Networks, June 2007, pp. 142 – 151
[3] M. Marina and S. R. Das, "On Demand Multipath Distance Vector Routing in Ad Hoc Networks," Proc. of the Int'l Conf. on Network Protocols(ICNP), December 2001, pp. 14-23.
[4] Holger Karl, Andreas Willig , "A short survey of wireless sensor networks," TKN Technical Report TKN-03-018, Berlin, October 2003.
[5] W.B. Heinzelman, A.P. Chandrakasan, H. Balakrishnan, "An application-specific protocol architecture for wireless microsensor networks," IEEE Transactions on Wireless Communications, Vol. 1, October 2002, pp. 660-670.
[6] E. Kranakis, H. Singh, and J. Urrutia, "Compass Routing on Geometric Networks," 11th Canadian Conference on Computational Geometry, August 1999.
[7] O. Younis, S. Fahmy, "HEED: A Hybrid, Energy-Efficient, Distributed Clustering Approach for Ad Hoc Sensor Networks," IEEE Trans on Mobile Computing, Vol. 3, No. 4, 2004.
[8] A.B. McDonald, T. Znati, "A mobility-based framework for adaptive clustering in wireless ad hoc networks. IEEE Journal on Selected Areas in Communications," special Issue on Wireless Ad Hoc Networks, vol. 17, no. 8, pp. 1466–1487, August 1999.
[9] S. Basagni, "Distributed clustering for ad hoc networks," Proceedings of the 1999 International Symposium on Parallel Architectures, Algorithms, and Networks.
[10] K. Imran Ali, M. Sajjad, S. Mubashir, M. Sana, "Efficient Overlay Multicast Routing for Hybrid Networks, Malaysian Journal of Computer Science, vol. 24, no. 2, pp: 84-97, 2011.
[11] A.P. Jardosh, E.M, Belding-Royer, K.C Almeroth, S. Suri, "Toward realistic mobility models for mobile ad hoc networks," Proc ACM MOBICOM, San Diego, CA, Sep. 2003, pp. 217–229.
[12] M. Saad, M. Sajjad, B. Kashif, K. Samee Ullah, "Stable Path Multi-Channel Routing With Extended Level Channel Assignment," International Journal of Communication Systems vol: 25, no. 7, pp:887-902.
[13] C. Tracy, B. Jeff, D.Vanessa, "A Survey of Mobility Models for Ad hoc Network Research. Special Issue on Mobile Ad hoc Networking: Research, Trends and Applications," Journal of Wireless Communications and Mobile Computing, 2002, pp. 483-502.
[14] H. Xiaoyan, G. Mario, P. Guangyu, C.A. Ching-Chuan, "Group Mobility Model for Ad HocWireless Networks," Proceedings of the 2nd ACM international workshop on Modeling, analysis and simulation of wireless and mobile systems.
[15] S. Giordano, I. Stojmenovic, "Position based routing in ad hoc networks, a taxonomy", Ad Hoc Wireless Networking, 2004, pp. 103-136.
[16] S. Basagni, "Distributed clustering for ad hoc networks," International Symposium on Parallel Architectures, Algorithms, and Networks.
[17] M. Yu, J. H. Li, R. Levy, "Mobility Resistant Clustering in Multi-Hop Wireless Networks," Journal OF NETWORKS, VOL. 1, NO. 1, pp. 12-19, May 2006.
[18] W. Heinzelman, A. Chandrakasan, H. Balakrishnan (2000), "Energy-Efficient Communication Protocol for Wireless Microsensor Networks," 33rd Hawaii International Conference on System Sciences, 2000.
[19] S. Giordano, I. Stojmenovic, "Position based ad hoc routes in ad hoc networks," Handbook of Ad Hoc Wireless Networks, CRC Press, 2003, Chapter 16, 1-14.
[20] Z. Jin, Y. Jian-Ping, Z. Si-Wang, L. Ya-Ping, L. Guang, "A Survey on Position-Based Routing Algorithms in Wireless Sensor Networks," Algorithms 2, No. 1, pp. 158-182.
[21] K. Atta Ur Rehman, M. Sajjad, H. Khizar, K. Samee Ullah, "Clustering-based power-controlled routing for mobile wireless sensor networks. International Journal of Communication Systems (IJCS)," vol. 4, no. 25, 2012, pp. 529-542, 2011.
[22] M. Sajjad, W. Daniel, M. Stefan, Position-based Routing Protocol for Low Power Wireless Sensor Networks. Journal of Universal Computer Science, vol. 16, no. 9, pp.1215-1233, 2010.
[23] T. Paczesny, D. Paczesny , J. Weremczuk, R. Jachowicz, "Position-based Routing Protocol for Low Power Wireless Sensor Networks," Journal of Universal Computer Science, vol. 16, no. 9, 2010, pp.1215-1233.
[24] S. Karim, A. Ahmed, "Geographic Protocols in Sensor Networks," Encyclopedia of Sensors, American Scientific Publishers (ASP), 2006.
[25] M. Liliana, C. Arboleda, N. Nasser, "Cluster-based routing protocol for mobile sensor networks," Proceedings of the 3rd international conference on Quality of service in heterogeneous wired/wireless networks, August 2006, Waterloo, Ontario, Canada.
[26] A. Shahzad, M. Sajjad, "Distributed Grid based Robust Clustering Protocol for Mobile Sensor Networks," International Arab Journal of Information Technology (IAJIT), vol.9, No.2, October 2011.
[27] A. Shahzad, M. Sajjad, "Distributed Efficient Multi hop Clustering Algorithm for Mobile Sensor Networks," International Arab Journal of Information Technology (IAJIT) vol.8, No.3, July 2011.
[28] R. Moses, D. Krishnamurthy, R. Patterson, "A self localization method for wireless sensor networks," EURASIP Journal on Applied Signal Processing, October 2003, pp.348-358.
[29] A. Varga, "The OMNeT++ Discrete Event Simulation System," Proc. European Simulation Multiconference (ESM 2001), Prague, Czech Republic, 2001.
[30] W. Drytkiewicz, S. Sroka, V. Handziski, A. Koepke, and H. Karl, "A Mobility Framework for OMNeT++," 3rd International OMNeT++, January 2003.